# Electric spectroscopy of vortex states and dynamics in magnetic disks

Minori Goto,[1*] Hiroshi Hata,[1] Akinobu Yamaguchi,[1,2 **] Yoshinobu Nakatani,[3] Takehiro Yamaoka,[4] Yukio Nozaki[1***] and Hideki Miyajima[1]

[1]*Department of Physics, Keio University, 3-14-1, Hiyoshi, Kohoku-ku, Yokohama, Kanagawa, 223-8522, Japan*

[2]*PRESTO, JST, Honcho, 4-1-8, Kawaguchi, Saitama, 332-0012, Japan*

[3]*University of Electro-Communications, 1-5-1, Chofugaoka, Chofu, Tokyo, 182-8585, Japan*

[4]*SII Nanotechnology Inc., RBM Tsukiji Bldg., Shintomi, Chuo, Tokyo, 104-0041, Japan*

Corresponding authors: *M.G. (E-mail: minori510@live.jp)

**A.Y. (E-mail: yamaguch@phys.keio.ac.jp, aki-yamaguchi@aist.go.jp)

***Y. N. (E-mail: nozaki@phys.keio.ac.jp)

**Abstract**

Spin-polarized radio frequency (RF) currents and RF-Oersted fields resonantly excite a magnetic vortex core confined in a micron-scale soft magnetic disk. In this study, we measured the rectifying voltage spectra caused by the anisotropic magnetoresistance oscillation due to the gyration of the vortex with different polarity and chirality. The measured spectra are presented such that we can determine the vortex properties and strength of the spin torques and Oersted field accurately and directly through analytical calculation.

## I. INTRODUCTION

Currently, there is significant interest in the physical properties of artificial mesoscopic magnets with a confined structure. While considerable attention has been devoted to transport properties, because of their applications in magnetic storage devices and magnetic sensors, the radio frequency (RF) response of these systems has also garnered interest. Studies of the ferromagnetic resonance (FMR) spectrum provide a rich source of information on fundamental magnetic properties, including the unique anisotropies and internal magnetization dynamics of mesoscopic magnets. The RF oscillation of magnetization has been recently observed in single patterned ferromagnetic film as well as in metallic multilayers; moreover, several investigations have been conducted to clarify the magnetic dynamics and realize potential applications such as microwave detectors and oscillators. In these systems, spin waves are considered responsible for phase locking microwave oscillators and for rectification of RF currents passed through ferromagnetic microwave guides.

In particular, the rectification provides highly sensitive detection of the information and dynamics of microscopic spin structure. The spin-polarized current passing through twisted magnetic structures such as giant magnetoresistnace devices and a magnetic domain wall gives rise to a torque on the local magnetization vectors. This torque is called the spin transfer torque, which originates from the transfer of spin angular momentum from the conduction electron spin to the local magnetization. This effect allows for direct and local manipulation of the magnetization and is a promising writing mechanism for new nonvolatile memories. The spin transfer torque originates from the conduction electron spin that follows the local magnetization through s-d interaction; however, the driving torque consists of not only the spin transfer torque but also the Oersted field. A precise measurement of the contribution of the torques is necessary to manipulate the magnetization. We find that the rectifying spectra can precisely detect the vortex dynamics. The measurement spectra allow us to determine the vortex properties as well as the strength of the driving torques accurately and directly through developed analytical calculation.

A vortex in a micron- or nanostructured magnetic thin-film element forms when the in-plane magnetization curls around a vortex core. The vortex core magnetization turns out of the plane to minimize the exchange energy.[1, 2] When a vortex confined in a disk is driven out of an equilibrium state by either a magnetic field or a spin-polarized current, the vortex core gyrates around its equilibrium position.[3-5] This soliton-like motion of the magnetic vortex attracts considerable attention in the physics of magnetic materials.

The magnetic vortex is characterized by two binary properties, a chirality (clockwise or counterclockwise direction ($C$ = ±1) of the in-plane rotating magnetization) and a polarity (up or down direction ($P$ = ±1) of the vortex core magnetization). As a result, there are four different ground states of the vortex.[6-8] Thus, an understanding of the stability and dynamical behavior of the magnetic vortex as well as a way to electrically detect the vortex states is a major requirement for developing magnetic data storage technology.

In this study, we develop a theory of vortex dynamics including the contributions of the spin torque and the Oersted field, and demonstrate the electrical detection of both the polarity and the chirality of the vortex in an $Fe_{19}Ni_{81}$ disk using a current-induced rectifying effect. [9]

This paper is organized as follows. In Sec. II, the analytical model is described. We develop the analytical model concerning the rectifying spectrum based on the magnetoresitance oscillation from the vortex core gyration through the simultaneous application of the static field and the RF current and field. The experimental setup is presented in Sec. III. Section IV describes the experimental results of the rectifying spectra measured in the disks. In addition, we explain the analysis of the rectifying spectra and the detection of the polarity and chirality of the vortex. We focus on their characteristic dependences on the direction of the static field and the RF current and field. In Sec. V, the conclusions are summarized.

## II. THEORETICAL MODEL

The vortex motion and rectifying voltage can be calculated using the following analytical model. First, the total energy in a magnetic disk is calculated in Sec. II-A. Second, the trajectory of the vortex core position $\boldsymbol{r}$ is derived in Sec. II-B. Third, the temporal variation of the anisotropic magnetoresistance (AMR) originated by the trajectory is calculated in Sec. II-C.

## II-A Potential energy

The energy $U$ in the magnetic disk consists of the Zeeman, magnetostatic, and exchange energies. In this section, we analytically calculate these energies.

We begin by calculating the Zeeman energy in the disk. The Zeeman energy of the vortex or anti-vortex confined in the disk was calculated by Krüger *et al.*[10,11] Based on the computation from this study, we calculated the Zeeman energy of the vortex in a circular disk using the approximation to the second-order of the vortex core position $\boldsymbol{r}$. The following model derives the magnetization $\boldsymbol{M}$ in the circular disk:

$$\begin{aligned}\boldsymbol{M} &= \boldsymbol{M}(\boldsymbol{x}_{\mathrm{d}} - \boldsymbol{r}) = \boldsymbol{M}(\boldsymbol{\rho}) \\ &= CM_{\mathrm{s}}\begin{pmatrix} \sin\varphi \\ -\cos\varphi \end{pmatrix} \\ &= CM_{\mathrm{s}}\begin{pmatrix} \dfrac{y_{\mathrm{d}} - y}{\sqrt{(x_{\mathrm{d}} - x)^2 + (y_{\mathrm{d}} - y)^2}} \\ -\dfrac{x_{\mathrm{d}} - x}{\sqrt{(x_{\mathrm{d}} - x)^2 + (y_{\mathrm{d}} - y)^2}} \end{pmatrix}.\end{aligned} \tag{1}$$

Here, the coordinate systems $\boldsymbol{x}_{\mathrm{d}} = (x_{\mathrm{d}}, y_{\mathrm{d}})$, $\boldsymbol{r} = (x, y)$, and $\boldsymbol{\rho} = \boldsymbol{x}_{\mathrm{d}} - \boldsymbol{r} = (\rho, \varphi)$ are defined as shown in Fig. 1. $\boldsymbol{x}_{\mathrm{d}}$ denotes the Cartesian coordinate system whose origin is the center of the disk, while $\boldsymbol{\rho}$ is defined by the coordinate system whose origin is the vortex core position. $\boldsymbol{r}$ is the vortex core position in the Cartesian coordinate system. $C$ and $M_{\mathrm{s}}$ are chirality and saturation magnetization, respectively. Variation of magnetization with $z$-direction was neglected to simplify the problem. In addition, we neglected the core magnetization because it is too small to influence the system energy.

Taking into account Eq. (1), the Zeeman energy in circular disk $U_Z$ is calculated as

$$
\begin{aligned}
U_Z &= -\int_{\text{disk}} \boldsymbol{M}\cdot\boldsymbol{H} d^3 x_{\text{d}} \\
&= -CM_{\text{s}}L\int_{\text{disk}} \frac{H_x(y_{\text{d}}-y)-H_y(x_{\text{d}}-x)}{\sqrt{(x_{\text{d}}-x)^2+(y_{\text{d}}-y)^2}} d^2 x_{\text{d}}
\end{aligned}, \tag{2}
$$

where $\boldsymbol{H} = (H_x , H_y)$ and $L$ are the external field and thickness of the disk, respectively. We calculated the Taylor series for the vortex core position $\boldsymbol{r}$ in Eq. (2) and truncated it after the second-order. Then, we obtain

$$
\begin{aligned}
U_Z &\cong -CM_{\text{s}}L\int_{\text{disk}} d^2 x_{\text{d}} \left\{ x\cdot\left(\frac{\partial}{\partial x}\frac{H_x(y_{\text{d}}-y)-H_y(x_{\text{d}}-x)}{\sqrt{(x_{\text{d}}-x)^2+(y_{\text{d}}-y)^2}}\right)\Bigg|_{x=0,y=0} \right. \\
&\qquad \left. + y\cdot\left(\frac{\partial}{\partial y}\frac{H_x(y_{\text{d}}-y)-H_y(x_{\text{d}}-x)}{\sqrt{(x_{\text{d}}-x)^2+(y_{\text{d}}-y)^2}}\right)\Bigg|_{x=0,y=0} \right\} \\
&= \pi CM_{\text{s}}Lr_{\text{disk}}(H_x y - H_y x) \\
&\equiv \boldsymbol{q}\cdot(\boldsymbol{H}\times\boldsymbol{r}),
\end{aligned} \tag{3}
$$

where

$$
\boldsymbol{q} = Cq_0\boldsymbol{e}_z = \pi CM_{\text{s}}Lr_{\text{disk}}\boldsymbol{e}_z \tag{4}
$$

defines the stiffness coefficient vector. Here, $r_{\text{disk}}$ is the radius of disk, and $\boldsymbol{e}_i$ ($i$ = $x$, $y$, $z$) indicates the unit vector. Using the Zeeman energy described by Eq. (3), the force on the vortex core is calculated as follows:

$$
\begin{aligned}
-\frac{\partial U_Z}{\partial \boldsymbol{r}} &= -\frac{\partial}{\partial \boldsymbol{r}}(\boldsymbol{q}\cdot(\boldsymbol{H}\times\boldsymbol{r})) \\
&= -\boldsymbol{q}\times\boldsymbol{H}
\end{aligned}. \tag{5}
$$

This result is in agreement with Krüger's result.

Next, to elucidate the calculation of the magnetostatic energy in the circular disk, we consider Taylor approximation for the vortex core position $\boldsymbol{r}$ and truncate Taylor series after the second-order. When the vortex core is driven by in-plane external field, the magnetic poles appear on the side of the disk shown in Fig. 1. The magnetic poles generate a demagnetizing field, and as a result, a restoring force is exerted on the vortex core. After calculating the demagnetizing field, we obtain the total magnetostatic energy in the disk for the magnetization distribution described by Eq. (1) and the demagnetizing field. The density of magnetic poles $M_{\mathrm{p}}$ on the side of the disk is calculated as

$$M_{\mathrm{p}}(\theta) = CM_{\mathrm{s}} \frac{(y_{\mathrm{d}} - y)\cos\theta - x_{\mathrm{d}}\sin\theta}{\sqrt{x_{\mathrm{d}}^2 + (y_{\mathrm{d}} - y)^2}} . \tag{6}$$

Here, $\theta$ indicates the angle with respect to the $x$-axis: $\theta = \tan^{-1}(y_{\mathrm{d}}/x_{\mathrm{d}})$. Considering only the case where the vortex core moves to the $y$-direction because of the rotational symmetry in the circular disk, the demagnetizing field $\boldsymbol{H}_d$ at the center of the disk is calculated as

$$\boldsymbol{H}_d = \int_{\partial(\mathrm{disk})} -\frac{M_{\mathrm{p}}(\theta')}{4\pi\mu_0} \frac{\boldsymbol{x}_{\mathrm{d}}'}{r_{\mathrm{disk}}^3} d^2 x_{\mathrm{d}}' . \tag{7}$$

Here, $\partial$(disk) indicates the side area of the disk as shown in Fig. 1. Using the calculation for the second-order of $\boldsymbol{r}$, the demagnetizing field described by Eq. (7) is approximately reduced to

$$\boldsymbol{H}_d \cong \frac{CM_{\mathrm{s}}L}{4\pi\mu_0 r_{\mathrm{disk}}^2} y\boldsymbol{e}_x . \tag{8}$$

By expanding Eq. (1), we derive the magnetization $\boldsymbol{M}$ to the first-order of the radius $r$ as follows:

$$\begin{aligned} \boldsymbol{M} \cong & \frac{CM_{\mathrm{s}}}{r_{\mathrm{d}}}(y_{\mathrm{d}}\boldsymbol{e}_x - x_{\mathrm{d}}\boldsymbol{e}_y) \\ & - \frac{CM_{\mathrm{s}}}{r_{\mathrm{d}}^3}(x_{\mathrm{d}}^2\boldsymbol{e}_x + x_{\mathrm{d}}y_{\mathrm{d}}\boldsymbol{e}_y)y \cdot \end{aligned} \tag{9}$$

Here, we define $r_{\mathrm{d}} = (x_{\mathrm{d}}^2 + y_{\mathrm{d}}^2)^{1/2}$. Substituting Eqs. (8) and (9) into the following equation:

$$U_m = -\int_{\mathrm{disk}} \frac{1}{2} \boldsymbol{M} \cdot \boldsymbol{H}_d d^3 x_{\mathrm{d}} , \tag{10}$$

we obtain

$$U_m = \frac{\pi M_s^{\ 2} L^2}{8\mu_0 r_{\text{disk}}} y^2 \equiv \frac{1}{2}\kappa_m y^2 \quad . \tag{11}$$

Here, we define the effective stiffness coefficient as the following:

$$\kappa_m = \frac{\pi M_s^{\ 2} L^2}{4\mu_0 r_{\text{disk}}} \quad . \tag{12}$$

Finally, we consider the exchange energy given by the gradient of the magnetization in Eq. (1). The exchange energy is given by

$$U_{\text{ex}} = \int_{\text{disk}} -\frac{A}{M_s^{\ 2}} \boldsymbol{M}\Delta\boldsymbol{M}\, d^3x_{\text{d}} = A\int_{\text{disk}} \frac{d^3x_{\text{d}}}{\rho^2} \quad , \tag{13}$$

where $A$ is the exchange stiffness coefficient. Unfortunately, the density of the exchange energy in Eq. (13) diverges to infinity at the center of the vortex core. Thus, we consider only the exchange energy contribution through the magnetic structure near the side of the disk. Let us consider and calculate the exchange energy in detail. Figure 2 shows the schematic image of the disk and magnetization. The gray area indicates the circular disk. The dotted area indicates the magnetization distribution; the center of this area is the vortex core. The exchange energy is equal to the integral in the disk (the gray area); however, this area includes the singular point (vortex core). Furthermore, this exchange energy is equal to (the integral after the vortex core shifts) – (the integral before it shifts). That is, it is equal to (the gray blank area) – (the white dotted area) in Fig. 2, because the gray dotted area is a common area between the area before and after the vortex core displaced. Therefore, the exchange energy is given by

$$
\begin{aligned}
U_{\mathrm{ex}} &= A\int_{\delta(\mathrm{disk})} \frac{d^3 x_{\mathrm{d}}}{\rho^2} \\
&= AL\int_{-r_{\mathrm{disk}}}^{r_{\mathrm{disk}}} dx_{\mathrm{d}} \left\{ \int_{-\sqrt{r_{\mathrm{disk}}^2 - x_{\mathrm{d}}^2}}^{-\sqrt{r_{\mathrm{disk}}^2 - x_{\mathrm{d}}^2}+y} dy_{\mathrm{d}} \frac{1}{x_{\mathrm{d}}^2 + (y_{\mathrm{d}} - y)^2} \right. \\
&\qquad \left. - \int_{\sqrt{r_{\mathrm{disk}}^2 - x_{\mathrm{d}}^2}}^{\sqrt{r_{\mathrm{disk}}^2 - x_{\mathrm{d}}^2}+y} dy_{\mathrm{d}} \frac{1}{x_{\mathrm{d}}^2 + (y_{\mathrm{d}} - y)^2} \right\} \\
&\cong -\frac{ALy^2}{r_{\mathrm{disk}}^2} \int_{-r_{\mathrm{disk}}}^{r_{\mathrm{disk}}} dx_{\mathrm{d}} \frac{2\sqrt{r_{\mathrm{disk}}^2 - x_{\mathrm{d}}^2}}{r_{\mathrm{disk}}^2} \\
&= -\frac{\pi AL}{r_{\mathrm{disk}}^2} y^2 \\
&\equiv -\frac{1}{2}\kappa_{\mathrm{ex}} y^2
\end{aligned} \tag{14}
$$

Here, δ(disk) indicates the (gray blank) − (white dotted) area in Fig. 2. We approximate the exchange energy with the calculation to the second-order of the vortex core position ***r***. Here, introducing the effective stiffness coefficient with

$$
\kappa_{\mathrm{ex}} = \frac{2\pi AL}{r_{\mathrm{disk}}^2}, \tag{15}
$$

we summarize the restoring force given by

$$
\begin{aligned}
U(\boldsymbol{r}) &= U_m + U_{ex} \\
&= \frac{1}{2}\kappa \boldsymbol{r}^2
\end{aligned} \tag{16}
$$

Here, the effective stiffness coefficient is defined as

$$
\begin{aligned}
\kappa &= \kappa_m - \kappa_{\mathrm{ex}} \\
&= \frac{\pi M_{\mathrm{s}}^2 L^2}{4\mu_0 r_{\mathrm{disk}}} - \frac{2\pi AL}{r_{\mathrm{disk}}^2}
\end{aligned} \tag{17}
$$

Equation (17) indicates that the restoring force that consists of the magnetostatic energy and the exchange energy is in a competitive relationship. In the micron-scale disk, the exchange force is negligibly smaller than the magnetostatic force. For $\kappa_m > \kappa_{ex}$, the vortex core is confined in the disk, while for $\kappa_m < \kappa_{ex}$, the vortex is driven out of the disk. At $\kappa_m = \kappa_{ex}$, the critical thickness $L_c$ for the boundary between the vortex state and single domain state is calculated by

$$L_c = \frac{8\mu_0 A}{M_s^{\,2} r_{disk}} . \quad (18)$$

Equation (18) corresponds to Hoffmann's result without the following factor:[12-14]

$$\frac{1}{\pi^2}\left\{\ln\left(\frac{r_{disk}}{a}+1\right)+\gamma_e\right\} \sim 0.86 . \quad (19)$$

Here, $a$ and $\gamma_e$ are the lattice constant (~3 Å in $Ni_{81}Fe_{19}$) and the Euler constant (~0.5772).[12] The inconsistency between our result Eq. (18) and Hoffmann's exact solution is due to the approximation of the energy to the second-order of the vortex core position $\boldsymbol{r}$.

Finally, combined with the Zeeman, magnetostatic, and exchange energies, the potential energy in the system is given by

$$U(\boldsymbol{r}) = \frac{1}{2}\kappa \boldsymbol{r}^2 + \boldsymbol{q}\cdot(\boldsymbol{H}\times\boldsymbol{r}) . \quad (20)$$

**II-B Position of the vortex core**

In this section, we examine the vortex dynamics in the circular disk under the application of the external static field $\boldsymbol{H}_{ext}$, RF spin torque $\boldsymbol{u}$, and RF-Oersted field $he^{i\omega t}\boldsymbol{e}_y$. For the analytical calculations, we begin with the modified Thiele equation including the spin torque term [15,16]

$$\boldsymbol{G}(P)\times(\boldsymbol{u}-\dot{\boldsymbol{r}})=-\frac{\delta U}{\delta \boldsymbol{r}}-\alpha D\dot{\boldsymbol{r}}+\beta D\boldsymbol{u} \qquad (21)$$

where $\boldsymbol{G}(P)=-PG_0\boldsymbol{e}_z=-P\frac{2\pi LM_{\mathrm{s}}}{\gamma}\boldsymbol{e}_z$ is the gyrovector. $\boldsymbol{u}$ is the spin torque interaction term of the current and the magnetization, which is given by $\boldsymbol{u}=u_0e^{i\omega t}\boldsymbol{e}_x=\frac{\mu_B p(-Je^{i\omega t})}{eM_{\mathrm{s}}}\boldsymbol{e}_x$ and is proportional to the current density $J$, Bohr magneton $\mu_{\mathrm{B}}$, and spin polarization ratio $p$ with electron charge $e$. Here, we assumed that the current is injected to the $x$-axis. $\alpha$ is the Gilbert damping constant, $\beta$ is the nonadiabatic contribution to the spin-transfer torque and $D$ is the diagonal element of damping tensor, which is given by $D=\frac{G_0}{2}\ln\left(\frac{r_{\mathrm{disk}}}{\xi}\right)$ with vortex core radius $\xi$.[17] $U$ denotes the potential energy given by Eq. (20) with stiffness coefficients described by Eqs. (4) and (17). The magnetic field $\boldsymbol{H}$ is given by $\boldsymbol{H}=\boldsymbol{H}_{\mathrm{ext}}+\mathrm{Re}\{he^{i\omega t}\boldsymbol{e}_y\}$. The external field $\boldsymbol{H}_{\mathrm{ext}}=(H_x, H_y)$ is applied in the disk plane. $h$ is the amplitude of RF-Oersted field generated by the RF current flowing through the disk and electrode.

We assumed the vortex core position to be written by the following form:[15]

$$\boldsymbol{r}=\begin{pmatrix}x\\ y\end{pmatrix}=\begin{pmatrix}x_0+\mathrm{Re}\{X\mathrm{e}^{i\omega t}\}\\ y_0+\mathrm{Re}\{Y\mathrm{e}^{i\omega t}\}\end{pmatrix}. \qquad (22)$$

Here, $x_0$ and $y_0$ are the equilibrium position of the vortex core in the static magnetic field. $X$ and $Y$ are the complex oscillation amplitudes. Substituting Eq. (22) into Eq. (21), we obtain the following equation:

$$i\omega\begin{pmatrix}\alpha D & -G_0P\\ G_0P & \alpha D\end{pmatrix}\begin{pmatrix}X\\ Y\end{pmatrix}\mathrm{e}^{i\omega t}+\kappa\begin{pmatrix}x_0+X\mathrm{e}^{i\omega t}\\ y_0+Y\mathrm{e}^{i\omega t}\end{pmatrix}=\begin{pmatrix}Cq_0(H_y+h\mathrm{e}^{i\omega t})+\beta Du_0\mathrm{e}^{i\omega t}\\ G_0Pu_0\mathrm{e}^{i\omega t}-Cq_0H_x\end{pmatrix}. \qquad (23)$$

For the zero RF current, namely $J$ = 0, all terms concerned with the oscillation that correspond to the RF

spin torque $\boldsymbol{u}$, RF-Oersted field $h$, and oscillation amplitude $\boldsymbol{X}$ are removed. Therefore, by setting $J = u_0 = h = X = Y = 0$ in Eq. (23), we obtain the equilibrium position:

$$\begin{pmatrix} x_0 \\ y_0 \end{pmatrix} = \frac{Cq_0}{\kappa}\begin{pmatrix} H_y \\ -H_x \end{pmatrix}. \tag{24}$$

Substituting Eq. (24) into Eq. (23), we obtain the new equation describing the vortex dynamics:

$$\begin{pmatrix} \kappa + i\omega\alpha D & -i\omega G_0 P \\ i\omega G_0 P & \kappa + i\omega\alpha D \end{pmatrix}\begin{pmatrix} X \\ Y \end{pmatrix} = \begin{pmatrix} Cq_0 h + \beta D u_0 \\ G_0 P u_0 \end{pmatrix}. \tag{25}$$

Calculating Eq. (25), we obtain the oscillation amplitude of vortex core:

$$\begin{pmatrix} X \\ Y \end{pmatrix} = \begin{pmatrix} X' + iX'' \\ P(Y' + iY'') \end{pmatrix} \tag{26}$$

with

$$\begin{aligned}
X'(\omega) &= u_0 \frac{\tilde{\beta}\tilde{\kappa}(\omega_\gamma{}^2 - \omega^2) + (1+\tilde{\alpha}\tilde{\beta})\alpha_*\omega^2}{(1+\tilde{\alpha}^2)\{(\omega_\gamma{}^2 - \omega^2)^2 + (\alpha_*\omega)^2\}} + h\frac{C\tilde{q}_0\{\tilde{\kappa}(\omega_\gamma{}^2 - \omega^2) + \alpha_*\tilde{\alpha}\omega^2\}}{(1+\tilde{\alpha}^2)\{(\omega_\gamma{}^2 - \omega^2)^2 + (\alpha_*\omega)^2\}} \\
X''(\omega) &= u_0\omega \frac{-\tilde{\beta}\tilde{\kappa}\alpha_* + (1+\tilde{\alpha}\tilde{\beta})(\omega_\gamma{}^2 - \omega^2)}{(1+\tilde{\alpha}^2)\{(\omega_\gamma{}^2 - \omega^2)^2 + (\alpha_*\omega)^2\}} + h\omega\frac{C\tilde{q}_0\{-\tilde{\kappa}\alpha_* + \tilde{\alpha}(\omega_\gamma{}^2 - \omega^2)\}}{(1+\tilde{\alpha}^2)\{(\omega_\gamma{}^2 - \omega^2)^2 + (\alpha_*\omega)^2\}} \\
Y'(\omega) &= u_0 \frac{\tilde{\kappa}(\omega_\gamma{}^2 - \omega^2) + (\tilde{\alpha} - \tilde{\beta})\alpha_*\omega^2}{(1+\tilde{\alpha}^2)\{(\omega_\gamma{}^2 - \omega^2)^2 + (\alpha_*\omega)^2\}} - h\frac{C\tilde{q}_0\alpha_*\omega^2}{(1+\tilde{\alpha}^2)\{(\omega_\gamma{}^2 - \omega^2)^2 + (\alpha_*\omega)^2\}} \\
Y''(\omega) &= u_0\omega \frac{-\tilde{\kappa}\alpha_* + (\tilde{\alpha} - \tilde{\beta})(\omega_\gamma{}^2 - \omega^2)}{(1+\tilde{\alpha}^2)\{(\omega_\gamma{}^2 - \omega^2)^2 + (\alpha_*\omega)^2\}} - h\omega\frac{C\tilde{q}_0(\omega_\gamma{}^2 - \omega^2)}{(1+\tilde{\alpha}^2)\{(\omega_\gamma{}^2 - \omega^2)^2 + (\alpha_*\omega)^2\}}
\end{aligned}. \tag{27}$$

Here, $X'$ ($X''$) and $Y'$ ($Y''$) are the $x$- and $y$-component of the real-part (imaginary-part) of the oscillation amplitude of the vortex core position, respectively. In addition, we introduce the following reduced parameters to simplify the notation:[16]

$$\tilde{\alpha} = \alpha \frac{D}{G_0},\ \tilde{\beta} = \beta \frac{D}{G_0},\ \tilde{\kappa} = \frac{\kappa}{G_0},\ \tilde{q}_0 = \frac{q_0}{G_0},\ \alpha_* = \frac{2\tilde{\alpha}\tilde{\kappa}}{1+\tilde{\alpha}^2},\ \omega_\gamma = \frac{\tilde{\kappa}}{\sqrt{1+\tilde{\alpha}^2}}. \quad (28)$$

As seen in Eq. (26), the polarity $P$ is included only in the $y$ component. This is because the polarity-dependent gyroforce $\boldsymbol{G}(P)\times\boldsymbol{u}$ in Eq. (21) points in the direction of the $y$-axis. Finally, the vortex core position is described by Eqs. (22), (24), (26), and (27).

## II-C Rectifying spectra

The electrical resistance arising from the AMR effect in the $Fe_{19}Ni_{81}$ disk depends on the position of the vortex core. When the direction of the electric current is parallel to the magnetization, the magnetoresistance increases. On the other hand, the magnetoresistance decreases when the direction of the current is perpendicular to the magnetization. Figure 3 shows the relationship between the vortex core position and the resistance. The resistance decreases with moving the vortex core in the direction of the current ($\pm x$ direction) as shown in Figs. 3(a) and (b), while it increases with moving the vortex core in the direction perpendicular to the current ($\pm y$ direction) as shown in Figs. 3(c) and (d). Thus, we approximate the disk resistance near the center of the disk by the following equation: [18]

$$R = R_0 - a_x\left(x - x_R\right)^2 + a_y\left(y - y_R\right)^2. \quad (29)$$

Here, $x$ and $y$ are the vortex core position. $x_R$ and $y_R$ are the center of the electrode gap, the detail of the electrode position is presented in Secs. III and IV. Proportionality constants are defined as $a_x$ and $a_y$. $R_0$ is the resistance for $x= x_R$ , $y= y_R$. Substituting Eq.(22) into Eq.(29), resistance $R$ is given by

$$
\begin{aligned}
R(t) = R_0 - a_x\{&(x_0 - x_R)^2 + X'^2\cos^2\omega t + X''^2\sin^2\omega t \\
&+ 2(x_0 - x_R)X'\cos\omega t - 2X'X''\cos\omega t\sin\omega t - 2(x_0 - x_R)X''\sin\omega t\} \\
+ a_y\{&(y_0 - y_R)^2 + Y'^2\cos^2\omega t + Y''^2\sin^2\omega t \\
&+ 2P(y_0 - y_R)Y'\cos\omega t - 2Y'Y''\cos\omega t\sin\omega t - 2P(y_0 - y_R)Y''\sin\omega t\}
\end{aligned} \quad . \qquad (30)
$$

Considering Ohm's law with RF current $I=I_0\exp(i\omega t)$, we calculate

$$
\begin{aligned}
V(t) &= I(t)R(t) \\
&= I_0R_0\cos\omega t - a_xI_0\{(x_0 - x_R)^2\cos\omega t + X'^2\cos^3\omega t + X''^2\cos\omega t\sin^2\omega t \\
&\qquad + 2(x_0 - x_R)X'\cos^2\omega t - 2X'X''\cos^2\omega t\sin\omega t - 2(x_0 - x_R)X''\cos\omega t\sin\omega t\} \\
&\quad + a_yI_0\{(y_0 - y_R)^2\cos\omega t + Y'^2\cos^3\omega t + Y''^2\cos\omega t\sin^2\omega t \\
&\qquad + 2P(y_0 - y_R)Y'\cos^2\omega t - 2Y'Y''\cos^2\omega t\sin\omega t - 2P(y_0 - y_R)Y''\cos\omega t\sin\omega t\}
\end{aligned} \quad . \qquad (31)
$$

Using the trigonometric formula, we derive the direct current (DC) component of the rectifying voltage as the following:

$$
\begin{aligned}
V_{\mathrm{dc}} &= (V(t))_{\mathrm{dc}} \\
&= -a_xI_0(x_0 - x_R)X' + a_yI_0P(y_0 - y_R)Y'
\end{aligned} \quad . \qquad (32)
$$

Finally, substituting Eqs. (24) and (27) into Eq. (32), we obtain

$$
\begin{aligned}
V_{\mathrm{dc}} = &\frac{I_0a}{(1+\tilde{\alpha}^2)\{(\omega_\gamma{}^2 - \omega^2)^2 + (\alpha_*\omega)^2\}} \times \\
&\Bigg\{ -\frac{\tilde{q}_0}{\tilde{\kappa}}CH_yu_0\{\tilde{\beta}\tilde{\kappa}(\omega_\gamma{}^2 - \omega^2) + \alpha_*(1+\tilde{\alpha}\tilde{\beta})\omega^2\} - \frac{\tilde{q}_0{}^2}{\tilde{\kappa}}H_yh\{\tilde{\kappa}(\omega_\gamma{}^2 - \omega^2) + \alpha_*\tilde{\alpha}\omega^2\} \\
&+ x_Ru_0\{\tilde{\beta}\tilde{\kappa}(\omega_\gamma{}^2 - \omega^2) + \alpha_*(1+\tilde{\alpha}\tilde{\beta})\omega^2\} + x_R\tilde{q}_0Ch\{\tilde{\kappa}(\omega_\gamma{}^2 - \omega^2) + \alpha_*\tilde{\alpha}\omega^2\} \\
&- y_RPu_0\{\tilde{\kappa}(\omega_\gamma{}^2 - \omega^2) + (\tilde{\alpha} - \tilde{\beta})\alpha_*\omega^2\} + y_R\tilde{q}_0CPh\alpha_*\omega^2 \\
&- \frac{\tilde{q}_0}{\tilde{\kappa}}CPH_xu_0\{\tilde{\kappa}(\omega_\gamma{}^2 - \omega^2) + (\tilde{\alpha} - \tilde{\beta})\alpha_*\omega^2\} + \frac{\tilde{q}_0{}^2}{\tilde{\kappa}}PH_xh\alpha_*\omega^2 \Bigg\}
\end{aligned} \quad . \qquad (33)
$$

Here, considering that the proportionality constant $a_x$ and $a_y$ are on the same value because of the rotational symmetry in the circular disk, we assume $a_x \sim a_y \equiv a$. To reproduce the rectifying spectra, we use the physical values described in Table 1, such as the injected power, radius, and thickness. The Oersted field induced by the RF current is calculated using the equivalent circuit model and finite element method that we coded. First, to estimate the Oersted field, we analytically calculated the distribution of current density in the electrode using parameter: RF power $W = 5.0\times10^{-5}$ W. Second, Oersted field is derived from Biot-Savart's law, reflecting the sample shape. The numerical grid size is $1 \times 1 \times 1\ \mathrm{nm}^3$. After calculation of the Oersted field and current density in the system, we estimate the respective force from the Oersted field $q_0 h$ and the spin torque $G_0 u_0$. In comparison, the ratio between the Oersted field and spin torque is

$$qh : Gu_0 \sim 8 : 1 . \tag{34}$$

This result indicates that the contribution of RF-Oersted field is more dominant than the spin torque in this system with micron-scale radius. Considering the fact that RF-Oersted field dominantly drives the vortex core dynamics in Eq. (33), we approximate the rectifying spectrum as the following form:

$$V_{\mathrm{dc}} = -\frac{I_0 a \tilde{q}_0{}^2 h(\omega_\gamma{}^2 - \omega^2)}{(1+\tilde{\alpha}^2)\{(\omega_\gamma{}^2 - \omega^2)^2 + (\alpha_* \omega)^2\}} H_y + \frac{I_0 a \frac{\tilde{q}_0{}^2}{\tilde{\kappa}} h \alpha_* \omega^2}{(1+\tilde{\alpha}^2)\{(\omega_\gamma{}^2 - \omega^2)^2 + (\alpha_* \omega)^2\}} P H_x . \tag{35}$$

The first term is proportional to the $y$-component of the external field. This spectrum shape correlates with the dispersion function. The second term is proportional to the polarity and the $x$-component of the external field. This spectrum shape corresponds to the Lorentzian function. According to Eq. (35), it is possible to detect the sign of the polarity by the application of the static field in the $x$-direction. For detection of chirality, we focus on the external field dependence of the spectrum amplitude. The asymmetric electrode structure enables us to detect chirality even if the contribution of the spin torque is negligible because the current $I_0$ and the proportionality constant of AMR $a$ are dependent on the chirality and electrode structure (the detail is explained in Sec. IV). Therefore, we can detect both polarity and chirality by using Eq. (35) and fabricating the system with the asymmetric electrode.

**III. EXPERIMENTAL SETUP**

Schematic images of the measurement circuit and the magnetic disk are shown in Fig. 4. An $Fe_{19}Ni_{81}$ disk with a diameter of 3 μm and a thickness of 30 nm was patterned directly onto a polished MgO substrate by means of electron-beam lithography and lift-off technique. We designed the circular disk with two tags on both sides; this shape enabled us to control the vortex polarity and chirality, as shown in Figs. 5(a) and (b). The polarity was controlled by the magnetic field of 5 kOe normal to the plane,[22] and the chirality was controlled by the in-plane saturation magnetic field along magnetization reversal process.[23] To control the polarity and chirality simultaneously, we obliquely applied the magnetic field of 5 kOe to the sample with a tilt angle $\theta = 10°$, as shown in Fig. 5(c).

The coplanar waveguide structure made from Au (80 nm)/Cr (5 nm) was connected to the disk, and the center conductive strip line was placed on the disk.[24, 25] Here, the center of the electrode gap shifted to the $+x$ direction, as shown in Fig. 6. A sinusoidal continuous wave RF current with a power of $5.0\times10^{-5}$ W was subsequently injected into the disk by a signal generator in the frequency range from 50 to 150 MHz. A rectified DC voltage between the electrodes was measured via a bias tee, which separates the DC and RF components of the current. The coordinate system used in this study is also shown in Fig. 3. An external field was applied in the range of −50 Oe to 50 Oe, and was inclined at an angle $\varphi$ with respect to the $x$-axis. Here, the Hall electrode was not used. All measurements were performed at room temperature. The sense of direction of the static field is defined as positive along the $+x$ ($+y$) direction. For clarity, the measured spectra are shifted vertically.

## IV. EXPERIMENTAL RESULTS AND DISCUSSION

Figure 7 shows the external field dependence of the rectifying voltage spectra at the angle $\varphi = 0°$ for ($P$, $C$) = (+1, −1) and (−1, −1), respectively. Above $H = +9$ Oe, the spectra for $P = +1$ are convex upward, while the spectra for $P = -1$ are convex downward. In contrast, below $H = -10$ Oe, the spectra for $P = +1$ are convex downward, while the spectra for $P = -1$ are convex upward. The result shown in Fig. 7 indicates that the sign of the rectifying spectra is dependent on the direction of the static field and polarity. To understand the physical mechanism, we focus on the amplitude of the spectra with respect to the sign of the polarity and field.

Figures 8(a) and (b) show the amplitude of the rectifying voltage at $P = +1$ and −1, respectively. The amplitude $V_{peak}$ is defined by sum of the maximum value $V_{max}$ and the minimum value $V_{min}$ of the rectifying voltage, namely $V_{peak} = V_{max}+V_{min}$. The amplitude is proportional to the external field $H$, when $H$ is small. Comparing the result shown in Fig. 8(a) with that in Fig. 8(b), we notice that the field dependence of the $V_{peak}$ for the vortex with ($P$, $C$) = (+1, −1) shows the opposite behavior with respect to the other vortex states with ($P$, $C$) = (−1, −1). The gradient depends on the polarity (solid line in Fig. 8). This result is in good agreement with the analytical result expected from the second term in Eq. (35). Therefore, the field dependence of the spectra depends on polarity; as a result, it enables us to know the sign of polarity.[18]

Next, to detect chirality, we measured the rectifying spectra in the field at an angle $\varphi$ = 90°. Figures 9(a) and (b) show the external field dependence of the rectifying voltage spectra for ($P$, $C$) = (−1, +1) and (−1, −1), respectively. As shown in Fig. 9(a), we were only able to detect the resonant signals from the vortex gyration above the field of $H$= −9 Oe. In contrast, the resonant signals were only observed below the field of $H$ = 9 Oe, as shown in Fig. 9(b). The asymmetry in the external field dependence is explained as follows. When the static field is applied to the direction of $\varphi$ = 90°, the vortex core position displaces in the $x$-direction. The rectifying spectra were only observed while the vortex core gyrates in the gap of the electrodes. When the vortex core displaces under the electrodes, the rectifying spectra disappear. When the center of the electrode gap is shifted to the $x$-direction from the center of the disk ($x_R$ > 0 in Eq. (29)), this asymmetrical electrode position had an effect on the asymmetrical external field dependence of the rectifying voltage in Fig. 9. If symmetric electrodes come into contact with the disk, the symmetric field dependence of the rectifying spectra can be observed. When the external field is applied to the +$y$ direction with clockwise chirality, the vortex core moves to the +$x$ direction and the rectifying voltage can be detected, as shown in Fig. 9(a). On the other hand, when the external field is applied to the +$y$ direction with the counterclockwise chirality, the vortex core moves to the $-x$ direction and the rectifying voltage vanished, as shown in Fig. 9(b) because the vortex core displaces under the electrode.

We estimate the amplitude of the Lorentzian function and dispersion function using the following fitting function:

$$y = \frac{2A(\omega_\gamma^{\ 2} - \omega^2) + B(\alpha_* / \omega_\gamma)\omega^2}{(\omega_\gamma^{\ 2} - \omega^2)^2 + (\alpha_* \omega)^2}. \tag{36}$$

Here, $A$ and $B$ are the contribution of the dispersion and Lorentzian functions. Figures 10(a) and (b) show the rectifying voltage and fitting function in the external field of +28 Oe for $\phi = 0°$ and 90°, respectively. As shown in Fig. 10(a), the ratio between $A$ and $B$ is $A : B = -1 : 75$ at $\phi = 0°$. This result indicates that the Lorentzian function is dominant at $\phi = 0°$. In contrast, at $\phi = 90°$, the ratio between $A$ and $B$ is $A : B = -11 : 1$, as shown in Fig. 10(b). This result indicates that the dispersion function has a strong influence on the spectrum shape at $\phi = 90°$. If the driving force of the Oersted field is negligibly small in comparison with the spin torque, we can approximate the rectifying spectrum using Eq. (33) via the following form:

$$V_{\mathrm{dc}} = -\frac{I_0 a_x \dfrac{\tilde{q}_0}{\tilde{\kappa}} \alpha_* u_0 \omega^2}{(1+\tilde{\alpha}^2)\{(\omega_\gamma^{\ 2} - \omega^2)^2 + (\alpha_* \omega)^2\}} CH_y - \frac{I_0 a_y \tilde{q}_0 u_0 (\omega_\gamma^{\ 2} - \omega^2)}{(1+\tilde{\alpha}^2)\{(\omega_\gamma^{\ 2} - \omega^2)^2 + (\alpha_* \omega)^2\}} CPH_x . \tag{37}$$

The first term is proportional to the chirality and the $y$-component of external field, whose spectrum shape is similar to the Lorentzian function. The second term is proportional to the chirality, polarity, and the $x$-component of the external field. This spectrum shape corresponds to the dispersion function. The spectrum shape in Eq. (37) is different from Eq. (35), and is not in agreement with the experiment. This result indicates that the driving force of the Oersted field is dominant. These fittings using Eq. (36) based on Eq.

(33) not only lead to the detection of the polarity and chirality but also are important in distinguishing the respective contribution of driving torques.

## V. CONCLUSION

This study offered a highly sensitive electric detection of the vortex core dynamics in an $Fe_{19}Ni_{81}$ disk with different polarity and chirality. We demonstrated the dependence of rectifying voltage spectra excited by the RF current on the external field for $\phi = 0°$ and $\phi = 90°$, respectively. The vortex polarity and chirality can be determined by both the spectrum shape and the field dependence of rectifying spectra. These experimental spectra are in good agreement with our analytical model. In addition, the curve fitting based on the analytical calculation enabled us to distinguish the individual contributions of the driving torques. Our experimental result provides a way to detect the vortex states by using current-induced rectifying effect. Our study has presented a new clue to help understand vortex dynamics and develop magnetic memory or logic devices concerned with the vortex core.

## VI. ACKNOWLEDGMENT

We would like to thank Y. Kasatani, Dr. K. Hosono and Prof. J. Shibata for their valuable discussions. This work is partly supported by JST CREST, MEXT Grants-in-Aid for Scientific Research in a Priority Area and a JSPS Grants-in-Aid for Scientific Research. A. Y. also acknowledges support from the JST PRESTO program.

**Table**

**Table 1**

Estimated values of the physical quantity set in our experiment.

## Figure captions

**Figure 1**

Schematic image of the coordinate system in the disk. $\boldsymbol{x}_\mathrm{d} = (x_\mathrm{d}, y_\mathrm{d})$ and $\boldsymbol{\rho} = \boldsymbol{x}_\mathrm{d} - \boldsymbol{r} = (\rho, \varphi)$ are the Cartesian coordinate system and polar coordinate system, respectively. The origin of the former system is the center of the disk, while that of the latter is the vortex core. $\boldsymbol{r}$ is the vortex core position in the Cartesian coordinate system. $\partial(\mathrm{disk})$ indicates the side area of the disk.

**Figure 2**

Schematic image of the disk and magnetization distribution. The gray area indicates the circular disk. The dotted area indicates the magnetization structure of the vortex; the center of this area is the vortex core.

**Figure 3**

Schematic images of the domain in the magnetic disk and electric current. The resistance in (a) and (b) are smaller than those in (c) and (d).

**Figure 4**

Schematic image of the measurement circuit and magnetic disk.

**Figure 5**

Schematic images of polarity- and chirality-controlled magnetic disks. (a) Polarity-control method for applying magnetic field of 5 kOe normal to the plane. (b) Chirality-control method for applying in-plane saturation magnetic field. Two tags of circular disk nucleate a chirality-controlled vortex core. (c) The method for applying in-plane and out-of-plane magnetic field simultaneously. We apply the magnetic field of 5 kOe to the sample on the brazen oblique foundation. The tilt angle $\theta$ is 10°.

**Figure 6**

Atomic force microscope (AFM) image of the disk and electrode. White dot indicates an outline of the magnetic disk.

**Figure 7**

Polarity and external field dependence of rectifying voltage spectra. The thick line (thin line) is the spectra for $P = +1$ ($-1$).

**Figure 8**

Amplitude of the rectifying voltage. (a) $P = +1$ (b) $P = -1$.

**Figure 9**

Chirality and external field dependence of rectifying voltage spectra. (a) $C = +1$ (b) $C = -1$.

**Figure 10**

Thin line indicates the rectifying voltage spectra at $H$ = 28 Oe. Thick line indicates the fitting function using Eq. (36). (a) The external field is applied to the $x$-axis. (b) The external field is applied to the $y$-axis.

**Table 1**

| **physical quantity** | **symbol** | **value** |
|---|---|---|
| power | $W$ | $5.0\times10^{-5}$ W |
| disk radius | $r_{\mathrm{disk}}$ | $1.5\times10^{-6}$ m |
| thickness | $L$ | $3\times10^{-8}$ m |
| electrical resistance | $R$ | $5\times10^{1}$ Ω |
| saturation magnetization | $M_{\mathrm{s}}$ | 1.2 T |
| vortex core radius | $\xi$ | 5 nm |
| Gilbert damping constant | $\alpha$ | 0.01 |
| non-adiabatic spin torque [19-21] | $\beta$ | 0.02 |
| resonant frequency | $\omega_{\gamma}/2\pi$ | $8\times10^{7}$ Hz |
| electrode shift ($x$-axis) | $x_R$ | $2.9\times10^{-7}$ m |
| electrode shift ($y$-axis) | $y_R$ | $2.6\times10^{-7}$ m |
| proportionality constant | $a$ | $3.6\times10^{10}$ Ω/m$^2$ |

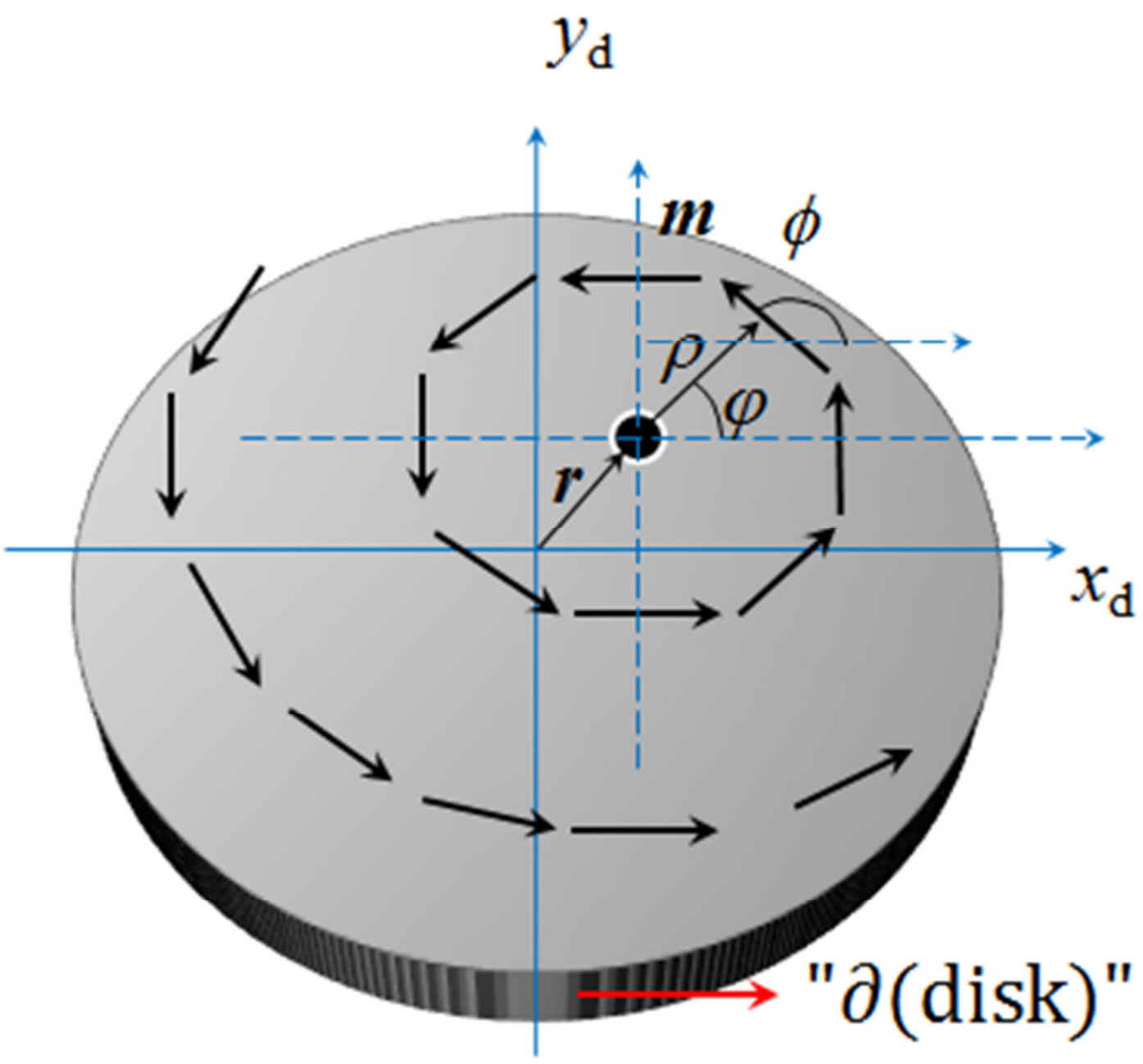

**Figure 1**

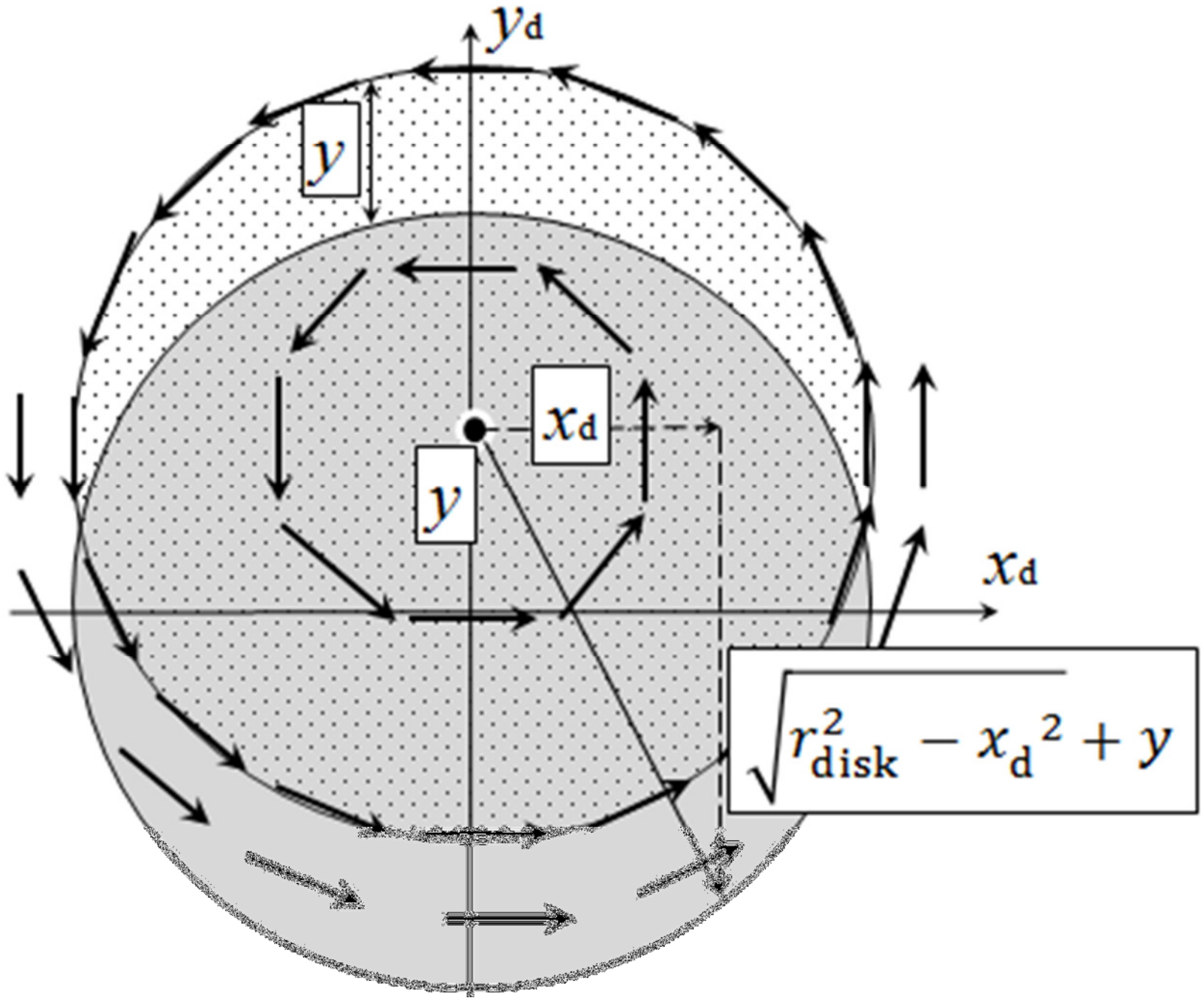

**Figure 2**

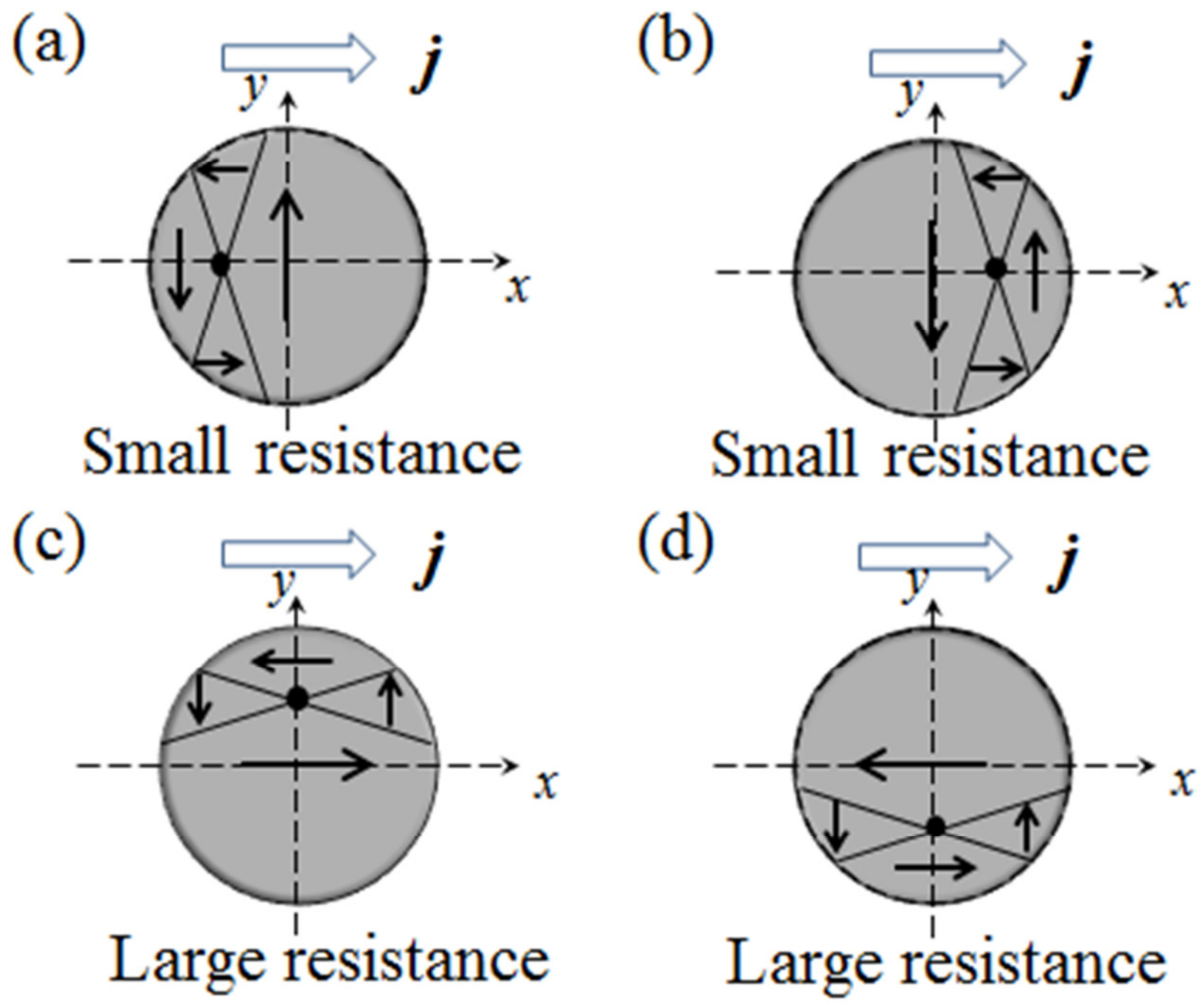

**Figure 3**

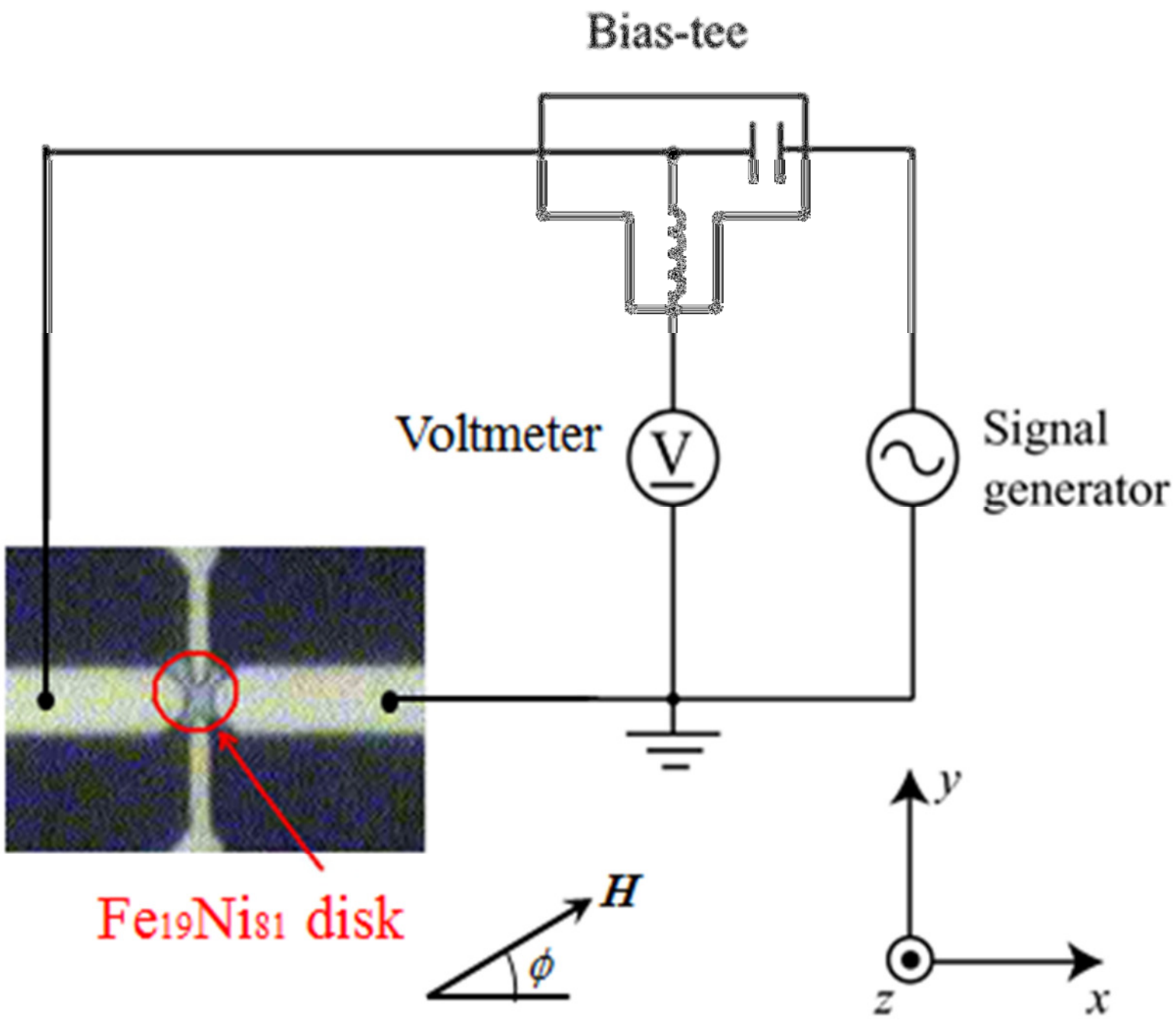

**Figure 4**

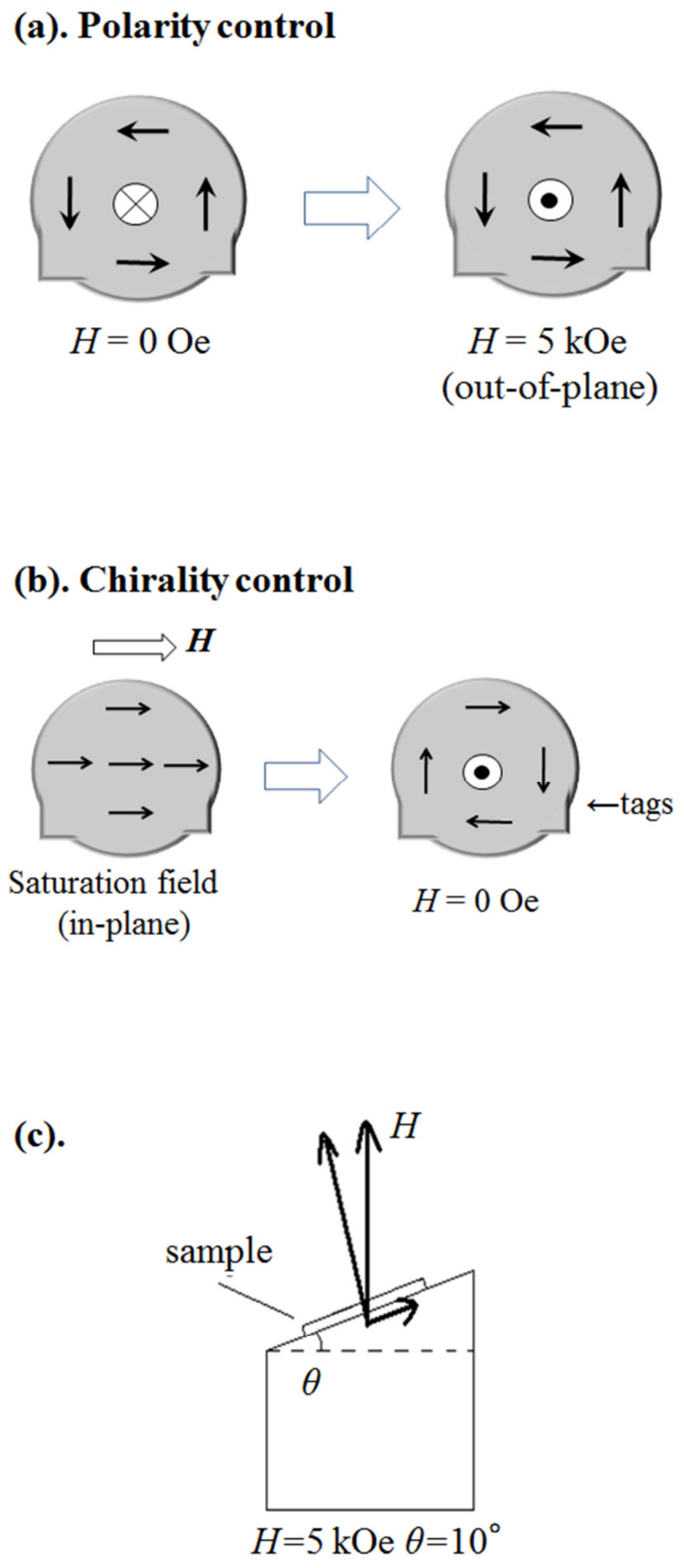

**Figure 5**

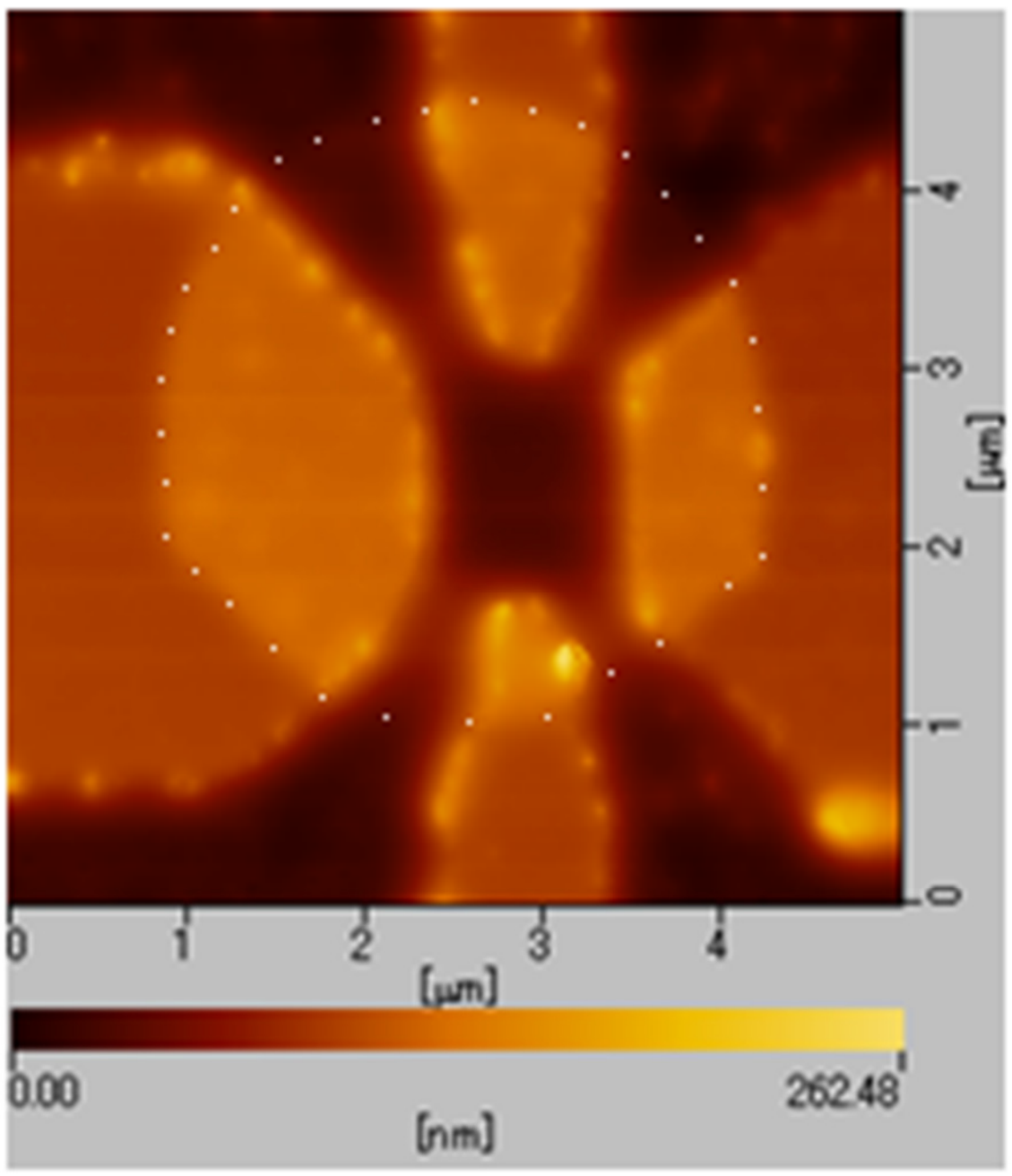

**Figure 6**

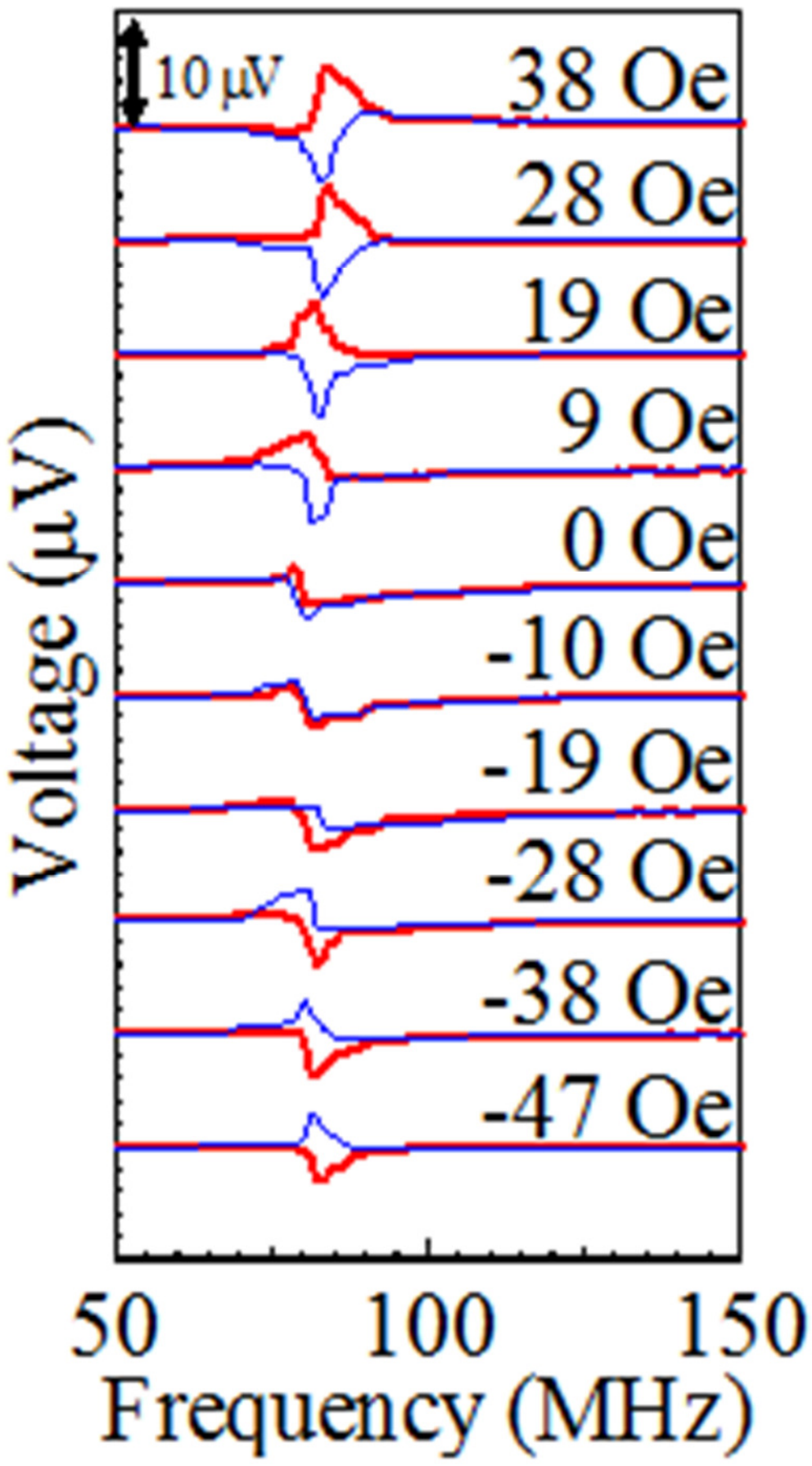

**Figure 7**

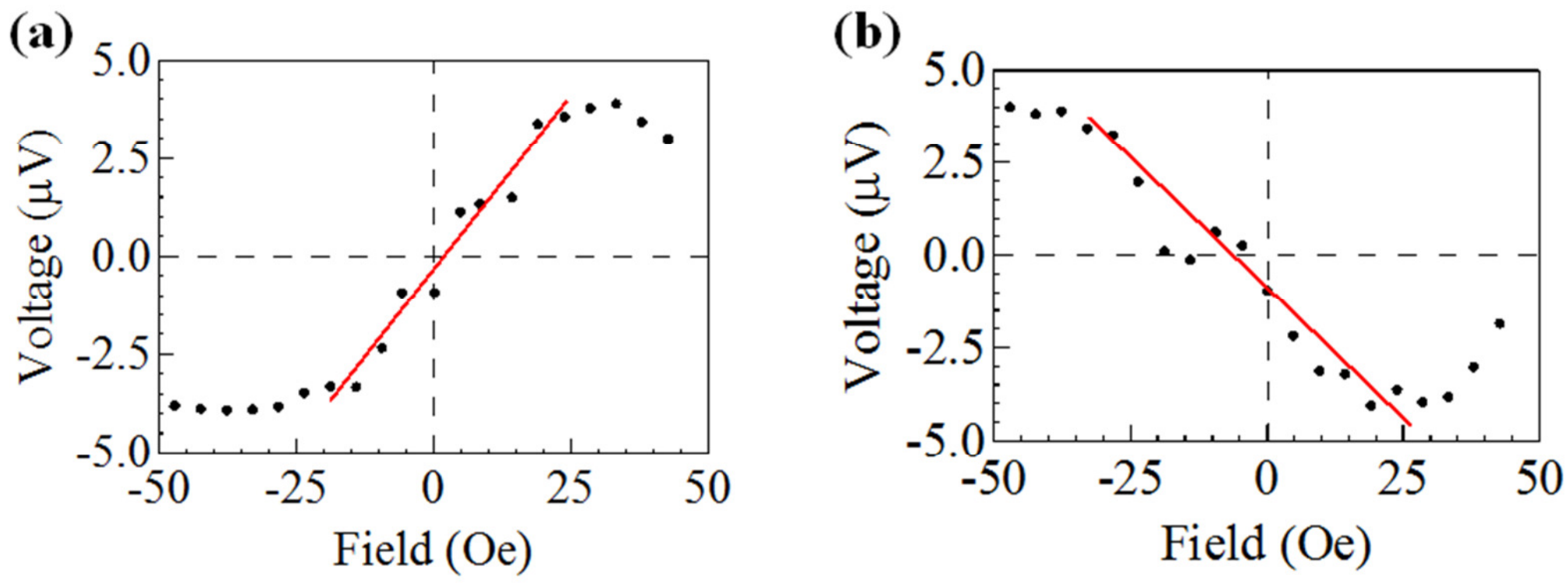

**Figure 8**

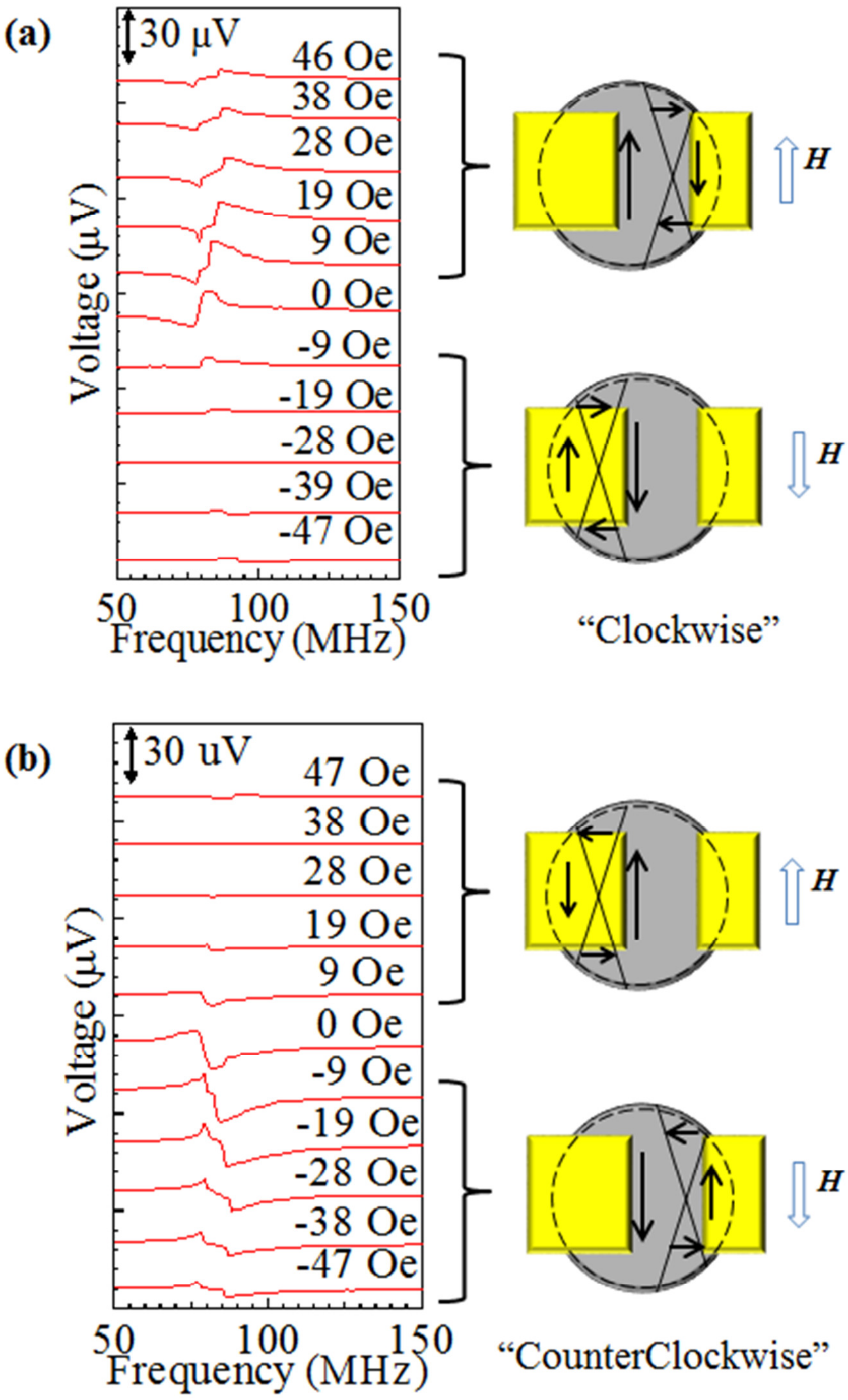

Figure 9

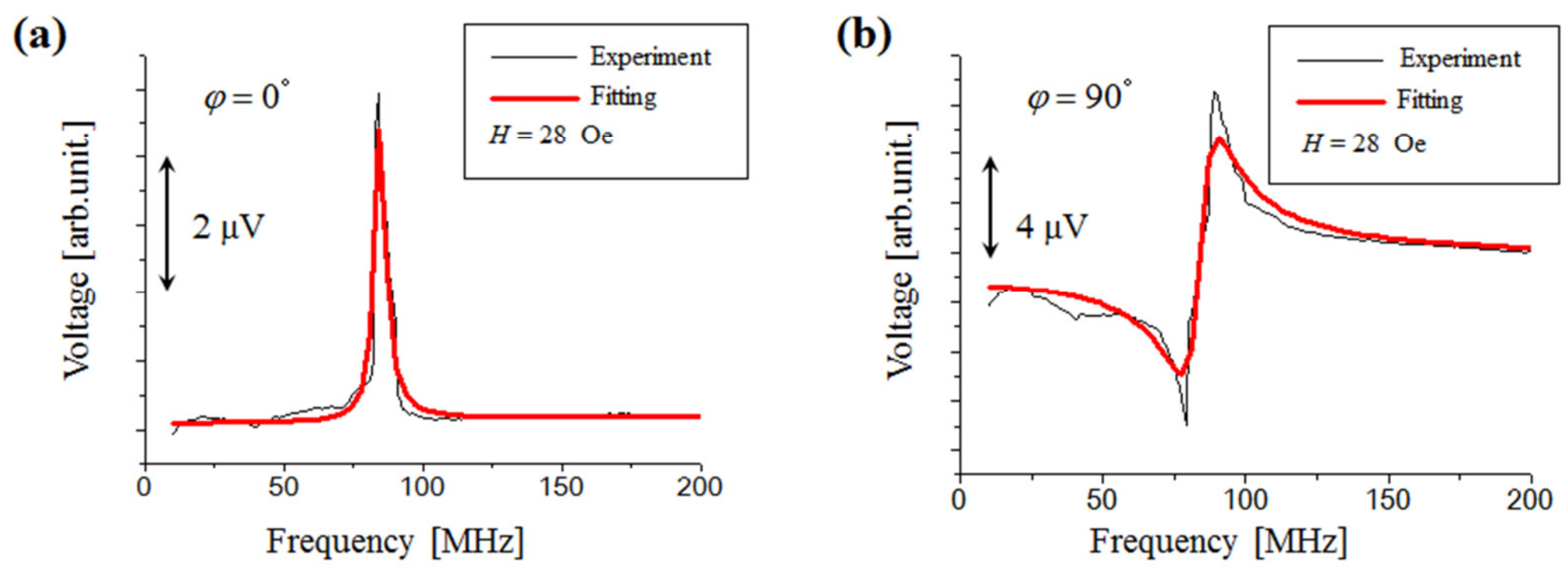

**Figure 10**